\documentclass[13pt]{article}
\usepackage{mathrsfs}
\usepackage{amsfonts}
\usepackage{amsmath,amssymb}
\begin{document}
\title{{\bf Remarks on the Sequential Products}\thanks{This project is supported by Zhejiang
Innovation Program for Graduates (YK2009002) and Natural Science
Foundation of China (10771191 and 10471124) and  Natural Science
Foundation of Zhejiang Province of China (Y6090105).}}
\author {{Liu Weihua,\,\,  Wu Zhaoqi,\,\, Wu Junde\thanks{Corresponding author E-mail: wjd@zju.edu.cn}}\\
{\small\it Department of Mathematics, Zhejiang University, Hangzhou
310027, P. R. China}}

\date{}
\maketitle

{\small\it \noindent {\bf Abstract}. In this paper, we show that
those sequential products which were proposed by Liu and Shen and Wu
in [J. Phys. A: Math. Theor. {\bf 42}, 185206 (2009), J. Phys. A:
Math. Theor. {\bf 42}, 345203 (2009)] are just unitary equivalent to
the sequential product $A\circ B=A^{\frac{1}{2}}BA^{\frac{1}{2}}$.

\vskip 0.1 in

}.

\vskip 0.1 in

{\bf Pacs.} {03.65.Aa, 03.65.Db}

\vskip 0.2 in

\noindent Quantum measurement theory is one of the key problems in
quantum theory, it contains a great many of mathematical problems
and philosophical problems. Also it has applications in quantum
information theory and quantum correction theory. The essential
difference between quantum measurement and classical measurement is
that the quantum measurement would make the system collapsed. It has
the follows four characteristics:

\noindent (1). Randomness. It is unpredictable and uncontrollable.

\noindent (2). Irreversibility. In general, measurement is
entropy-increasing procedure.

\noindent (3). Decoherence. Eliminate all the coherence of the
original state.

\noindent (4). Nonlocality. The collapse of the wave function is
nonlocal.

\vskip0.1in

\noindent In history, Heinsenberg, von Neumann, Birkhoff published
some important far-reaching fundamental works. In order to state our
main results, now, we need to recall some elementary notations.

\vskip0.1in

\noindent  Let $\cal L$ be a quantum-mechanical system and it be
represented by a complex Hilbert space $H$. Each self-adjoint
operator $A$ on $H$ satisfies that $0\leq A\leq I$ is said to be a
quantum effect ([1-2]). Quantum effects represent yes-no
measurements that may be unsharp. The set of quantum effects on $H$
is denoted by ${\cal E}(H)$. The subset ${\cal P}(H)$ of ${\cal
E}(H)$ consisting of orthogonal projection operators represents
sharp yes-no measurements. Let ${\cal T}(H)$ be the set of trace
class operators on $H$ and ${\cal S}(H)$ the set of density
operators on $H$, i.e., the state set of quantum system $\cal L$.

\vskip0.1in

As we knew, a quantum measurement can be described as a quantum
operation which is a completely positive linear mapping $\Phi: {\cal
T}(H)\rightarrow {\cal T}(H)$ such that for each $T\in {\cal S}(H)$,
$0\leq tr[\Phi (T)]\leq 1$ ([3-5]). For each $P\in {\cal P}(H)$, the
so-called L\"{u}ders operation $\Phi_L^P$ is defined by
$T\rightarrow PTP$, in physics, it implied that if the
quantum-mechanical system $\cal L$ is in state $W\in {\cal S}(H)$,
then the probability that the measurement $P$ is observed is given
by $p_{W}(P)=tr(PWP)$, moreover, the resulting state after the
measurement $P$ is observed is $\frac{PWP}{tr(PWP)}$ whenever
$tr(PWP)\neq 0$ ([4]). Each quantum effect $B\in {\cal E}(H)$ gives
 to a general L\"{u}ders operation $\Phi_L^B: T\rightarrow
B^{\frac{1}{2}}TB^{\frac{1}{2}}$. If $A, B\in {\cal E}(H)$, then the
composition operation $\Phi_L^B\circ \Phi_L^A$ defines a new
operation and is called a sequential operation as it is obtained by
performing first $\Phi_L^A$ and then $\Phi_L^B$. It is easily to
prove that $\Phi_L^B\circ
\Phi_L^A=\Phi_L^{A^{\frac{1}{2}}BA^{\frac{1}{2}}}$ $([5,
P_{26-27}])$. Let us denote $A^{\frac{1}{2}}BA^{\frac{1}{2}}$ by
$A\circ B$, then $A\circ B\in {\cal E}(H)$ and $\circ$ has the
following important properties ([6-7]):

\vskip0.1in

\noindent (S1). The map $B\rightarrow A\circ B$ is additive for each
$A\in {\cal E}(H)$, that is, if $B+C\leq I$,

then $(A\circ B)+(A\circ C)\leq I$ and $(A\circ B)+(A\circ C)=A\circ
(B+C)$.

\noindent (S2). $I\circ A=A$ for all $A\in {\cal E}(H)$.

\noindent (S3). If $A\circ B=0$, then $A\circ B=B\circ A$.

\noindent (S4). If $A\circ B=B\circ A$, then $A\circ
(I-B)=(I-B)\circ A$ and $A\circ (B\circ C)=(A\circ B)\circ C$

for all $C\in {\cal E}(H)$.

\noindent (S5). If $C\circ A=A\circ C$, $C\circ B=B\circ C$, then
$C\circ (A\circ B)=(A\circ B)\circ C$

and $C\circ (A+B)=(A+B)\circ C$ whenever $A+B\leq I$.

\vskip0.1in

Professor Gudder called  $A\circ B$ the sequential product of $A$
and $B$, it represents the quantum effect produced by first
measuring $A$ then measuring $B$ ([6-7]). In [8], Gudder asked: is
$A\circ B=A^{\frac{1}{2}}BA^{\frac{1}{2}}$ the only operation on
${\cal E}(H)$ which satisfies the properties (S1)-(S5) ? In [9], Liu
and Wu showed that if $H$ is a finite dimensional complex Hilbert
space, $f_z(u)$ is the complex-valued function defined on $[0, 1]$,
where $f_z(u)=\exp z(\ln u)$ if $u\in (0, 1]$ and $f_z(0)=0$, and
denote $A^i=f_i(A),\,\, A^{-i}=f_{-i}(A)$, then $A\circ_1
B=A^{1/2}A^{i}BA^{-i}A^{1/2}$ defined a new sequential product which
satisfies the properties (S1)-(S5), thus, Gudder's problem was
answered negatively.

\vskip0.1in

Note that the sequential product $A\circ
B=A^{\frac{1}{2}}BA^{\frac{1}{2}}=A^{\frac{1}{2}}B(A^{\frac{1}{2}})^*$
of $A$ and $B$ can only describe the instantaneous measurement, that
is, the measurement $B$ is completed at once after the measurement
$A$ is performed. In order to describe a more complicated process
where we allow a duration between the measurement $A$ with the
measurement $B$, then we need to replace $A^{\frac{1}{2}}$ with
$f(A)$, $ (A^{\frac{1}{2}})^*$ with $(f(A))^*$, where $f(A)$ is a
function of $A$ which describe the change of $A$ was made by the
duration between $B$ with $A$. Thus, we need to consider the
following general sequential product $f(A)B(f(A))^*$.

\vskip0.1in

By the above motivation, in [10], Shen and Wu proved the following
result:

\vskip0.1in

{\bf Theorem 1}. Let $H$ be a finite dimensional complex Hilbert
space, for each $A\in {\cal E}(H)$, $sp(A)$ the spectra of $A$ and
${\cal B}(sp(A))$ the set of all bounded complex Borel functions on
$sp(A)$. Take a $f_A\in {\cal B}(sp(A))$. Define $A\diamond
B=f_A(A)B(f_A(A))^*$ for $B\in{\cal E}(H)$. Then $\diamond$ has the
properties (S1)-(S5) iff the set $\{f_A\}_{A\in{\cal E}(H)}$
satisfies the following conditions:

\vskip 0.1 in

(i) $ \ $  For every $A\in{\cal E}(H)$ and $t\in sp(A)$, $|f_A(t)|=
\sqrt{t}$;

(ii) $ \ $ For any $A,B\in{\cal E}(H)$, if $AB=BA$, then there
exists a complex constant $\xi$ such that $|\xi|=1$ and
$f_A(A)f_B(B)=\xi f_{AB}(AB)$.

\vskip 0.1 in

Note that for each $A\in {\cal E}(H)$, we can take many $f_A\in
{\cal B}(sp(A))$ satisfies the conditions (i) and (ii), so, Theorem
1 told us that for each given finite dimensional complex Hilbert
space $H$, there are many sequential products on $({\cal
E}(H),0,I,\oplus)$.

\vskip 0.1 in

In this note, we show that these sequential products are unitary
equivalent to the sequential product $A\circ
B=A^{\frac{1}{2}}BA^{\frac{1}{2}}$.

\vskip 0.1 in

Firstly, we need the following:

\vskip 0.1 in

{\bf Lemma 1.1 ([10])}. If $\{f_A\}_{A\in{\cal E}(H)}$ satisfies the
conditions (i) and (ii) of Theorem 1, then we have

\vskip 0.1 in

(1) $ \ $ $f_A(A)\overline{f_A}(A)=\overline{f_A}(A)f_A(A)=A$,
$(f_A(A))^*=\overline{f_A}(A)$.

(2) $ \ $ If $0\in sp(A)$, then $f_A(0)=0$.

(3) $ \ $ If $A=\sum\limits^{n}_{k=1}\lambda_{k}E_{k}$, where
$\{E_{k}\}^{n}_{k=1}$ are pairwise orthogonal projections and
$\lambda_{k}\neq 0$, then
$f_A(A)=\sum\limits^{n}_{k=1}f_A(\lambda_{k})E_{k}$.

\vskip 0.1 in

Our main result is:

\vskip 0.1 in

{\bf Theorem 2}. Let $H$ be a finite dimensional complex Hilbert
space. Then the sequential product $f_A(A)B(f_A(A))^*$ on $({\cal
E}(H),0,I,\oplus)$ is unitary equivalent to the sequential product
$A\circ B=A^{\frac{1}{2}}BA^{\frac{1}{2}}$.

\vskip 0.1 in

{\bf Proof}. Let $A=\sum\limits^{n}_{k=1}\lambda_{k}E_{k}$ be the
spectra decomposition of $A$, where $\{E_{k}\}^{n}_{k=1}$ be
pairwise orthogonal projection operators and $\lambda_{k}>0,$ $k=1,
2, \cdots, n$. By condition (i) of Theorem 1, we have
$|f_A(\lambda_{k})|=\sqrt{\lambda_{k}}$, so
$f_A(\lambda_{k})=\sqrt{\lambda_{k}}e^{i\theta_k}$ for some real
number $\theta$. Let $E_0=I-\sum_{k=1}^nE_{k}$ and
$U=\sum_{k=1}^ne^{i\theta_k}E_{k}+E_0$. Then $U$ is an unitary
operator and it is easy to see that $AU=UA$, so by Lemma 1.1, we
have $f_A(A)=A^{1/2}U$. Thus, $A\diamond
B=f_A(A)B\overline{f_A}(A)=A^{1/2}UB(A^{1/2}U)^{*}=U(A^{1/2}BA^{1/2})U^{*}=U(A\circ
B)U^{*}$ and the conclusion is proved.

\vskip0.2in

\centerline{\bf References}

\vskip0.2in

\noindent [1] G. Ludwig, {\it Foundations of Quantum Mechanics
(I-II)},  Springer, New York, 1983

\noindent [2] G. Ludwig, {\it An Axiomatic Basis for Quantum
Mechanics (II)}, Springer, New York, 1086

\noindent [3] K. Kraus, {\it Effects and Operations},
Springer-Verlag, Beilin, 1983

\noindent [4] E. B. Davies, {\it Quantum Theory of Open Systems},
Academic Press, London, 1976

\noindent [5] P. Busch, M. Grabowski M and P. J. Lahti,  {\it
Operational Quantum Physics}, Springer-Verlag, Beijing Word
Publishing Corporation, 1999

\noindent [6] S. Gudder, G. Nagy, J. Math. Phys. {\bf 42}, 5212
(2001)

\noindent [7] S. Gudder, R. Greechie. Rep. Math. Phys. {\bf 49}, 87
(2002)

\noindent [8] S. Gudder, Inter. J. Theory. Phys. {\bf 44}, 2219
(2005)

\noindent [9] Liu W. H., Wu J. D. J. Phys. A: Math. Theor. {\bf 42},
185206 (2009)

\noindent [10] Shen J., Wu J. D., J. Phys. A: Math. Theor. {\bf 42},
345203 (2009)

\end{document}